\documentstyle[12pt]{article}
\def\beq{\begin{equation}}
\def\eeq{\end{equation}}
\def\bea{\begin{eqnarray}}
\def\eea{\end{eqnarray}}
\def\bq{\begin{quote}}
\def\eq{\end{quote}}

\def\gappeq{\mathrel{\rlap {\raise.5ex\hbox{$>$}}
{\lower.5ex\hbox{$\sim$}}}}

\def\lappeq{\mathrel{\rlap{\raise.5ex\hbox{$<$}}
{\lower.5ex\hbox{$\sim$}}}}

\begin{document}
\topmargin -0.5cm
\oddsidemargin -0.3cm
\pagestyle{empty}
\begin{flushright}
{HIP-1998-67/TH}
\end{flushright}
\vspace*{5mm}
\begin{center}
{\bf Canonical structure of Yang-Mills theory} \\
\vspace*{1.5cm} 
{\bf Christofer Cronstr\"{o}m}$^{*)}$ \\
\vspace{0.3cm}
Physics Department, Theoretical Physics Division  \\
FIN-00014 University of Helsinki, Finland \\
\vspace{0.5cm}
 
\vspace*{3cm}  
{\bf ABSTRACT} \\ \end{center}
\vspace*{5mm}
\noindent
I consider the problem of defining canonical coordinates and momenta in pure Yang-Mills theory, 
under the condition that Gauss' law is identically satisifed. This involves among other things 
particular boundary conditions for certain dependent variables. These boundary conditions are not 
postulated a priori, but arise as consistency conditions related to the equations of motion.
It is shown that the theory indeed has a canonical structure, provided one uses a special gauge 
condition, which is a  natural generalisation to Yang-Mills theory of the Coulomb gauge condition 
in electrodynamics. The canonical variables and Hamiltonian are explicitly constructed. 
Quantisation of the theory is briefly discussed.

\vspace*{1cm} 
\noindent
 
%11.15.-q, 04.20.Fy, 04.60.Ds 

\vspace*{2cm} 
\noindent 
$^{*)}$ e-mail address: Christofer.Cronstrom@Helsinki.fi.
\vspace*{1cm}
\begin{flushleft}
September 27, 1998
\end{flushleft}
\vfill\eject

%\pagestyle{empty}
%\clearpage\mbox{}\clearpage

\setcounter{page}{1}
\pagestyle{plain}

\section{Introduction}

In this paper I present a novel solution to an old problem in Yang-Mills theory \cite{YM}, namely
the problem of formulating the Yang-Mills equations of motion as a set of canonical Hamiltonian
equations. The canonical formulation of Yang-Mills theory has been considered as a neccessary
prerequisite for the development of a consistent quantisation procedure of this theory, as emphasised
e.g. in the textbook on gauge theory by Faddeev and Slavnov \cite{FaddeSlav}. I refer to this textbook
for an account of the state of the art in Yang-Mills theory around 1980. An account of more recent 
developments can be found e.g. in the textbook by Weinberg \cite{Weinberg}.

The early attempts to quantise Yang-Mills theory \cite{Schw 1}, \cite{Schw 2} were based on the
use of the Coulomb gauge condition, familiar from electrodynamics, in Yang-Mills theory. However,
as shown by Gribov \cite{Gribov} in 1977, the Coulomb gauge is not a proper gauge condition in
Yang-Mills theory. It should be noted here that the problems pointed out by Gribov concerning
the use of the Coulomb condition, the so-called Gribov ambiguities, do not really matter in 
perturbative calculations. 

Another possible approach to a canonical formalism in Yang-Mills theory \cite{Jackiw}, is based on 
the so-called Weyl gauge \cite{Weyl}, $A_{0} = 0$. However, in this formulation one simply disregards
the Gauss law, and considers only the canonical Hamiltonian formulation of the remaining Yang-Mills 
equations, which is a rather simple problem. Gauss' law is then in the quantised version of this theory
introduced by hand, as a condition on the states in the theory. While this may be an acceptable
approach to quantum Yang-Mills theory, it has, to the best of my knowledge, yielded only partial insight
into the structure  of pure Yang-Mills theory.

There has recently been an upsurge in interest in non-perturbative aspects of Yang-Mills theory, as
witnessed e.g. by a paper by Faddeev and Niemi \cite{FaddeNiemi}. It is emphasised by Faddeev and 
Niemi in their paper, that we are still lacking theoretical understanding of low-energy Yang-Mills 
theory, i.e. the non-perturbative regime, which is relevant for studying questions such as colour 
confinement. The problem of confinement is definitely a question which calls for a proper quantum 
Yang-Mills theory. A proper canonical formulation of the (semiclassical) Yang-Mills theory  may well be
a necessary prerequisite for this, or at least an appropriate starting point.  

The real confinement problem concerns the situation when quark fields are included in the formalism,
but this extension is not an essential difficulty in the formulation presented here. 

The canonical formulation developed in this paper differs from the Weyl gauge formulation in that 
Gauss' law is implemented identically (in principle). This has necessitated the introduction of a new
gauge condition \cite{ChrgenCg}, which is a straightforward and natural generalisation to Yang-Mills theory
of the ordinary Coulomb gauge condition in electrodynamics. The resulting canonical Yang-Mills theory, 
in its Schr\"{o}dinger quantised form,  resembles the corresponding theory based on the Weyl gauge in 
many ways, but is apparently not isomorphic to that theory.

\section{Notation and Conventions}  % Produces section heading.  Lower-level
                                    % sections are begun with similar 
                                    % \subsection and \subsubsection commands.

The basic variables in Yang-Mills theory \cite{YM} are the gauge field $G_{\mu \nu}$
and gauge potential $A_{\mu}$, respectively. These quantities take values in  a convenient 
(Hermitian) matrix representation  of  the Lie algebra of the gauge group G. The Lie algebra is 
defined by the structure constants $f_{ab}^{~~c}$ in the  commutator algebra of the Hermitian 
matrix representatives $T_{a}$ of the Lie algebra generators,
\begin{equation}
[T_{a}, T_{b}] = if_{ab}^{~~c}T_{c},
\label{eq:liealg}
\end{equation}
where appropriate summation over repeated Lie algebra indices $(a, b, c, ..., f)$ is understood.
We assume $G$  to be semisimple and compact. Then a  positive definite Lie algebra metric 
can be given in terms of the following Killing form $(h_{ab})$,
\begin{equation}
h_{ab} = -f_{ab'}^{~~c'}f_{bc'}^{~~b'}
\label{eq	:Killing}
\end{equation}
The inner product $(A, B)$ of any two Lie algebra valued quantites $A = A^{a}T_{a}$ and
$B = B^{a}T_{a}$ is defined as follows,
\begin{equation}
(A, B) = h_{ab}A^{a}B^{b}
\label{eq:inprod}
\end{equation}
The form $h_{ab}$ and its inverse $h^{ab}$ are used to lower and raise Lie algebra indices, 
respectively. 

In the notation introduced so far, we write the gauge potential $A_{\mu}$ as follows,
\begin{equation}
A_{\mu}(x) = A_{\mu}^{a}(x) T_{a}
\label{eq:potA}
\end{equation}
where the argument $x$ stands for a space-time point in Minkowski space. The gauge field 
$G_{\mu\nu}(A)$ is then given as follows,
\begin{equation}
G_{\mu\nu}(A) = \partial_{\nu}A_{\mu}(x) - \partial_{\mu}A_{\nu}(x) - ig[A_{\mu}(x), A_{\nu}(x)],
\label{eq:Gfield}
\end{equation}
where $g$ is an arbitrary nonvanishing real parameter, which is introduced for convenience.
An alternative to the notation (\ref{eq:Gfield}) is the following,
\begin{equation}
G_{\mu\nu}^{a}(A) = \partial_{\nu}A_{\mu}^{a}(x) - \partial_{\mu}A_{\nu}^{a}(x) + gf_{bc}^{~~a}
A_{\mu}^{b}(x), A_{\nu}^{c}(x)
\label{eq:Gindfield}
\end{equation}

The remaining notation is fairly conventional. Greek letters $\mu, \nu, ...$ are spacetime indices 
which take values in the range $(0,1,2,3)$. These indices are lowered (raised) with the standard 
diagonal Minkowski metric $g_{\mu\nu}\; (g^{\mu\nu})$ with signature $(+,-,-,-)$. Latin indices 
from the middle of the alphabet $(k, \ell, ...)$ are used as space indices in the range $(1, 2, 3)$. 
Unless otherwise stated, repeated indices are always summed over, be they Lie algebra-, spacetime- 
or space indices.

I finally state the gauge transformation formulae for the gauge potential and field, respectively,
in my notation. Let $\Omega(x)$ denote a general gauge transformation, the generic form of which is 
a unitary matrix of the following form,
\begin{equation}
\Omega(x) = \exp(ig\alpha^{a}(x) T_{a}),
\label{eq:gaugetrans}
\end{equation}
where the functions $\alpha^{a}(x)$ are any sufficiently smooth real-valued  functions on Minkowski 
space. The relevant gauge transformation formulae are as follows,
\begin{equation}
A_{\mu}(x)  \stackrel{\Omega}{\longrightarrow} A'_{\mu}(x) = \Omega^{-1}(x) A_{\mu}(x) \Omega(x)
+ \frac{i}{g} (\partial_{\mu}\Omega^{-1}(x))\Omega(x),
\label{eq:gtrA}
\end{equation}
and
\begin{equation}
G_{\mu\nu}(A) \stackrel{\Omega}{\longrightarrow} G_{\mu\nu}(A') = \Omega^{-1}(x) G_{\mu\nu}(A) \Omega(x)
\label{eq:gtrG}
\end{equation}

\section{The equations of motion and Gauss' law}

The Yang-Mills action $S$, with a general potential $A$, which at this stage is not 
assumed to satisfy any particular gauge condition, is as follows,
\begin{equation}
S = - \; \frac{1}{4}\; \int d^{4}x (G_{\mu\nu}(A), G^{\mu\nu}(A))
\label{eq:action}
\end{equation}
It is convenient to write the action (\ref {eq:action}) in terms of a Lagrangian $L$, although this
treats time- and space indices on an unequal footing. The action $S$ is the integral of a Lagrangian
$L$ in an appropriate time interval $[x^{0}_{i}, x^{0}_{f}]$. Thus,
\begin{equation}
S = \int_{x^{0}_{i}}^{x^{0}_{f}} dx^{0} L,
\label{eq:defSL}
\end{equation}
where 
\begin{equation}
L = -\frac{1}{2} \int_{V} d^{3} {\bf x}\left (G_{0k}(A), G^{0k}(A) \right ) - \frac{1}{4} \int_{V} d^{3} {\bf x}\left (G_{k\ell}(A), G^{k\ell}(A) \right ).
\label{eq:defL}
\end{equation}
In the expression (\ref {eq:defL}) for the Lagrangian $L$, the quantity $V$ is some appropriate domain
in ${\bf R}^{3}$, which yet has to be specified.  
 
As is well known, requiring the action (\ref{eq:action}) to be stationary with respect to local 
variations of {\em all} the potential components $A_{\mu}$, considered as independent
quantities, yields the follwing field equations,
\begin{equation}
\nabla_{\nu}(A) G^{\mu\nu}(A) = 0\;, \mu =  0,1,2,3.
\label{eq:fieldeq}
\end{equation}
The "covariant gradient" $\nabla_{\mu}(A)$ used above in Eq. (\ref {eq:fieldeq}) is a convenient 
notion,      
\begin{equation}
\nabla_{\mu}(A) \equiv \partial_{\mu} + ig\,[A_{\mu}, \;\;\;],
\label{eq:covgrad}
\end{equation}
which will be frequently used in what follows.

A perhaps more interesting system of equations would be obtained by coupling the gauge field to 
appropriate matter fields. Then the right hand side of Eq. (\ref{eq:fieldeq}) would be 
replaced by a covariantly conserved matter current. Such an addition is not absolutely essential 
for the questions pursued in this paper, and will therefore not be contemplated further here.

The non-Abelian Gauss law is obtained from the equations (\ref {eq:fieldeq}) for $\mu = 0$. 
Expressed in terms of the potential $A$ the non-Abelian Gauss law reads as follows, 
\begin{equation}
\nabla_{k}(A)\nabla^{k}(A) A^{0} -  \nabla_{k}(A){\dot A}^{k} = 0
\label{eq:nAGauss}
\end{equation}
Considered as an equation which determines the Lie algebra valued potential component $A_{0}$,
for given space components ${\bf A} = (A^{1}, A^{2}, A^{3})$, the equation 
(\ref {eq:nAGauss}) is a system of {\em linear, elliptic} partial differential equations with the
time variable $x^{0} \in [x^{0}_{i}, x^{0}_{f}]$ acting as a parameter. 

From now on it will be convenient to denote the time derivative of any quantity with a dot on top of
that quantity, thus for example
\begin{equation}
\dot{A}_{k}(x) \equiv \partial_{0}A_{k}(x)
\label{eq:dot}
\end{equation}

The apparent equations of motion  are obtained from the field equations (\ref {eq:fieldeq}) for 
$\mu \equiv k = 1,2,3$. These equations read as follows, separated into two groups of first-order
equations for convenience,
\begin{equation}
\dot{A}^{k} = \nabla^{k}(A) A^{0} - G^{0k}(A)\;, \dot{G}^{0k}(A)  = \nabla_{\ell}(A) G^{k\ell}(A) - ig [ A_{0},  G^{0k}(A) ] .
\label{eq:equmo}
\end{equation}

I then return to Eq. (\ref {eq:nAGauss}), i.e. the Gauss' law constraint. In order to analyse the 
solvability of Eq. (\ref {eq:nAGauss}), one has to specify the {\em domain} $V$ of the independent 
variables ${\bf x} = (x^{1}, x^{2}. x^{3}) \in {\bf R}^{3}$ in this equation. 

In what follows, I assume that the domain $V$ is a finite simply connected domain in ${\bf R}^{3}$
with a smooth boundary $\partial V$. Eventually one has to contemplate the situation where $V$ is 
allowed to grow indefinitely, but for the time being I consider only finite $V$. An example of a 
possible domain $V$ would be a ball $B_{R}$ of radius $R$, centered at the origin of space coordinates,
i.e.
\begin{equation}
B_{R}  = \left \{ {\bf x} | \mid {\bf x} \mid < R \right \},
\label{eq:domain}
\end{equation}
keeping in mind the possibility of the limit  $R \rightarrow \infty$ at some appropriate stage in the 
development of the formalism.

I will postpone till later the discussion of the boundary conditions which guarantee the {\em uniqueness} of a 
solution $A_{0}$ to Eq. (\ref {eq:nAGauss}), assuming that a solution actually
{\em exists}. The boundary conditions are  related to the canonical structure of the theory, 
as will be seen presently. 

The solution $A_{0}$ of Eq. (\ref {eq:nAGauss}), assuming the existence of a unique solution, is, in 
general, an $x$-dependent {\em functional} of {\em both} the space components ${\bf A}$ and their 
time derivatives ${\dot{{\bf A}}}$, i.e.
\begin{equation}
A_{0} = A_{0}\left \{{\bf A}, {\dot{{\bf A}}}\right \}
\label{eq:Aofunct}
\end{equation}

The question is then whether the Yang-Mills system, which is originally defined by the action 
(\ref {eq:action}), permits a canonical structure when the potential component $A_{0}$ is a 
solution to Gauss' law (\ref {eq:nAGauss}), i.e. when $A_{0}$ is given in terms of ${\bf A}$ and 
${\dot{{\bf A}}}$ by the expression (\ref {eq:Aofunct}). This question can be analysed without 
specifying the actual functional form of the relation (\ref {eq:Aofunct}) in minute detail.

The Lagrangian of the Yang-Mills system, when the potential component $A_{0}$, is given by the 
relation (\ref {eq:Aofunct}), is  obtained simply by inserting the solution (\ref {eq:Aofunct}) 
for $A_{0}$ into the Lagrangian (\ref {eq:defL}) above. The resulting Lagrangian which will be 
called $L_{0}$, is then explicitly given as follows,
\begin{eqnarray}
\label{eq:L_{0}}
L_{0} & = & - \frac{1}{2}\int_{V} d^{3} {\bf x} \left (\nabla_{k}( A) A_{0}\left \{{\bf A}, \dot{{\bf A}}\right \}
 - \dot{{\bf A}}_{k}, \nabla^{k}( A) A^{0}\left \{{\bf A}, \dot{{\bf A}}\right \} - \dot{{\bf A}}^{k} \right ) \\
& & - \frac{1}{4} \int_{V} d^{3} {\bf x} \left ( G_{kl}(A), G^{kl}(A) \right )  \nonumber
\end{eqnarray}
At this point it is appropriate to check whether the action principle involving the Lagrangian $L_{0}$
in (\ref {eq:L_{0}}) above reproduces the equations of motion (\ref {eq:equmo}). It is perfectly
straightforward to verify the following result,
\begin{eqnarray}
\label{eq:varL_{0}}
\delta\;\int_{x^{0}_{i}}^{x^{0}_{f}} dx^{0} L_{0} & = & - \int_{x^{0}_{i}}^{x^{0}_{f}} dx^{0}
\int_{V}d^{3}{\bf x} \left (\delta A_{k}, \nabla_{0}(A)(\nabla^{k}(A) A^{0}\left \{{\bf A}, \dot{{\bf A}}\right \} - \dot{{\bf A}}^{k}) - \nabla_{\ell}G^{k\ell}(A) \right ) \\
& & - \int_{x^{0}_{i}}^{x^{0}_{f}} dx^{0} \int_{\partial V} d^{2}\sigma_{k} \left (\delta A^{0}\left \{{\bf A}, \dot{{\bf A}}\right \}, \nabla^{k}(A)A^{0}\left \{{\bf A}, \dot{{\bf A}}\right \} - \dot{{\bf A}}^{k} \right ) \nonumber
\end{eqnarray}
Here the boundary conditions for the solution $A_{0}$ to Gauss' law, i.e. the
system of linear elliptic partial differential equations (\ref {eq:nAGauss}), come into play. The 
surface term in Eq. (\ref {eq:varL_{0}}), if non-vanishing, destroys the variational principle which 
is supposed to lead to the equations of motion (\ref {eq:equmo}) with $A_{0}$ given by 
(\ref {eq:Aofunct}). It is therefore necessary to expurgate this surface term from the expression 
(\ref {eq:varL_{0}}). This can be done by declaring that the system of linear elliptic partial 
differential equations (\ref {eq:nAGauss}) for $A_{0}$ has to be solved with {\em fixed boundary values}
at $\partial V$, which results in the condition,
\begin{equation}
\delta A_{0}\left \{{\bf A}, \dot{{\bf A}}\right \}|_{{\bf x} \in \partial V}\; = 0
\label{eq:bc}
\end{equation}
for all admissible variations of the independent generalised coordinates ${\bf A}$ and velocities
$\dot{{\bf A}}$, respectively. From now on it will be understood that fixed boundary values, i.e.
boundary conditions of the Dirichlet type, will have to be used on the boundary $\partial V$
in solving the Gauss' law  (\ref {eq:nAGauss}). The appropriate Dirichlet boundary conditions will 
be discussed in  detail in a separate paper \cite{Chris4}. 

Under these circumstances the variational principle
\begin{equation}
\delta\;\int_{x^{0}_{i}}^{x^{0}_{f}} dx^{0} L_{0} = 0,
\label{eq:var2}
\end{equation}
leads to the following equations of motion,
\begin{equation}
\nabla_{0}(A)\left \{\nabla^{k}(A) A^{0}\left \{{\bf A}, \dot{{\bf A}}\right \} - \dot{{\bf A}}^{k}\right \} - \nabla_{\ell}G^{k\ell}(A) = 0,
\label{eq:eqmo2}
\end{equation}
as is evident from the relation (\ref {eq:varL_{0}}). Needless to say, the equations (\ref {eq:eqmo2})
are nothing but the original equations of motion (\ref {eq:equmo}), albeit in a slightly disguised 
form, with $A_{0}$ given by the solution (\ref {eq:Aofunct}) to Gauss' law (\ref {eq:nAGauss}).    
   
\section{Canonical formulation of the equations of motion}

\subsection{First attempt at constructing canonical momenta}

Using the Lagrangian (\ref {eq:L_{0}}) as a starting point, one may now attempt to construct
a canonical description of the system defined by the equations of motion (\ref {eq:eqmo2}). 

The formal definition of the canonical momentum $P_{k}^{a}$ conjugate to the coordinate 
$A_{a}^{k}$ is,
\begin{equation}
P_{k}^{a}(x^{0}, {\bf x}) \equiv \frac{\delta L_{0}}{\delta \dot{A}_{a}^{k}(x^{0}, {\bf x})} = \left ( \nabla_{k}(A)A_{0}\left \{{\bf A}, {\dot{{\bf A}}}\right \} \right )^{a}(x^{0}, {\bf x}) - \dot {A}_{k}^{a}(x^{0}, {\bf x}),
\label {eq:canmom1}
\end{equation}
where the condition (\ref {eq:bc}) has been used in the calculation of the functional derivative
of $L_{0}$ in (\ref {eq:canmom1}) above. In view of the fact that $A_{0}$ in Eq. (\ref {eq:canmom1})
satisfies Gauss' law (\ref {eq:nAGauss}), one finds immediately from (\ref {eq:canmom1}) that
\begin{equation}
\nabla^{k}(A)P_{k}(x^{0}, {\bf x}) \equiv 0.
\label{eq:divP=0}
\end{equation}
Now one is supposed to be able to solve Eq. (\ref {eq:canmom1}) for the generalised velocity
$\dot{A}_{k}^{a}$ in terms of ${\bf A}$ and ${\bf P}$, respectively. But this is impossible, 
since, Eq. (\ref {eq:canmom1}) can not be solved for the quantity $\Gamma$ defined below,
\begin{equation}
\Gamma(x^{0}, {\bf x}) \equiv \nabla_{k}(A)\dot{A}^{k}(x^{0}, {\bf x}),
\label{eq:defGamma}
\end{equation}
i.e. if one tries to derive an equation for the quantity $\Gamma$ defined above, from Eq. (\ref {eq:canmom1}),
one gets a completely vacuous identity for this quantity, as a result of Eq. (\ref {eq:divP=0}).

It would seem then, that the equations of motion (\ref {eq:eqmo2}) can not be written as a system
of canonical first order equations at all. However this is actually not the case. 

It turns out, that the quantity $\Gamma$ defined above in (\ref {eq:defGamma}), can be set to zero
as a consequence of the freedom of choosing a gauge in Yang-Mills theory. To be precise, 
the condition
\begin{equation}
\Gamma(x^{0}, {\bf x}) = 0
\label{eq:genCg}
\end{equation}
has recently been shown to be a {\em proper gauge condition}, called the generalised Coulomb gauge 
condition \cite{ChrgenCg}.

In what follows, it will be assumed that the generalised Coulomb gauge condition (\ref {eq:genCg})
is in force. Since the generalised velocities $\dot{A}^{k}$ then no longer are independent 
quantities, one cannot use the formula (\ref {eq:canmom1}) as it stands for the construction
of the canonical momentum variables. An alternative procedure that leads to a canonical 
formalism will be given below.

\subsection{Canonical coordinates, momenta and Hamiltonian}

It is now appropriate to summarise the various conditions on the generalised coordinates ${\bf A}$
and genralised velocities $\dot{{\bf A}}$, respectively, that have come out of the previous analysis.

In the first place, it is assumed, that the {\em generalised Coulomb gauge condition} (\ref {eq:genCg})
is in force, i.e. that
\begin{equation}
\nabla_{k}(A)\dot{A}^{k}(x^{0}, {\bf x}) = 0
\label{eq:genCgA}
\end{equation}
Then {\em Gauss' law} (\ref {eq:nAGauss}) simplifies, and takes the following form,
\begin{equation}
\nabla_{k}(A)\nabla^{k}(A) A^{0} = 0\;, {\bf x} \in V \subset {\bf R}^{3}
\label{eq:Gauss2}
\end{equation}
Hence {\em the solution} $A_{0}$ to Gauss' law (\ref {eq:Gauss2}) is a functional of the generalised
coordinates ${\bf A}$ {\em only},
\begin{equation}
A_{0} = A_{0}\left \{{\bf A}\right \}.
\label{eq:AofunctA}
\end{equation}
Fixed boundary conditions, independent of $\dot{{\bf A}}$, at the boundary $\partial V$ of the 
domain $V$ in Eq. (\ref {eq:Gauss2}) are used to define the solution (\ref {eq:AofunctA}) of the
present form of Gauss' law (\ref {eq:Gauss2}). As a consequence of this requirement the following 
condition will always be valid,
\begin{equation}
\delta A_{0}\left \{{\bf A}\right \}|_{{\bf x} \in \partial V}\; = 0
\label{eq:bc2}
\end{equation} 
for any admissible variations of the generalised coordinates ${\bf A}$.

Finally, the Lagrangian for the system under the conditons detailed above is recorded here for the 
sake of completeness. This Lagrangian, denoted by $L_{00}$, is nothing but the Lagrangian of Eq.
(\ref {eq:L_{0}}), but with the solution (\ref {eq:AofunctA}) for $A_{0}$ in stead of 
(\ref {eq:Aofunct}). Explicitly,
\begin{eqnarray}
\label{eq:L_{00}}
L_{00} & = & - \frac{1}{2}\int_{V} d^{3} {\bf x} \left (\nabla_{k}( A) A_{0}\left \{{\bf A}\right \}
 - \dot{{\bf A}}_{k}, \nabla^{k}( A) A^{0}\left \{{\bf A}\right \} - \dot{{\bf A}}^{k} \right ) \\
& & - \frac{1}{4} \int_{V} d^{3} {\bf x} \left ( G_{kl}(A), G^{kl}(A) \right )  \nonumber
\end{eqnarray}

I will now implement the gauge conditions (\ref {eq:genCgA}) as a {\em constraint}, by means of a
Lie algebra, or rather matrix valued Lagrange multiplier field $C(x)$, which is used to modify the 
Lagrangian $L_{00}$ in (\ref {eq:L_{00}}) in an appropriate way,
\begin{equation}
L_{00} \rightarrow L' = L_{00} + \int_{V} d^{3}{\bf x} (C(x), \nabla_{k}(A)\dot{A}^{k})
\label{eq:Lconstr}
\end{equation}
The Lagrange multiplier technique as such is well known in general as well as in the  
particular case of  gauge systems \cite{HennTeitel}. 

One can now make a direct transition to a Hamiltonian using the modified Lagrangian 
(\ref {eq:Lconstr}) above, in a manner described in the general case by Berezin \cite{Berezin}. 
The starting point
is the familiar definition of canonical momentum variables $\pi_{k}^{a}$,
\begin{equation}
\pi_{k}^{a}(x^{0}, {\bf x}) \equiv \frac{\delta L'}{\delta \dot{A}_{a}^{k}(x^{0}, {\bf x})} = \left (\nabla_{k}(A) A_{0}\left \{{\bf A} \right \} \right )^{a} - \dot{A}_{k}^{a} - \left ( \nabla_{k}(A) C \right )^{a},
\label{eq:canpi}
\end{equation}
where $x^{0}$ is a fixed parameter with variable ${\bf x} \in V$. The argument $(x^{0}, {\bf x})$
is omitted in the last term in (\ref {eq:canpi}) above. Similar simplifications of notation will be
used freely below, whenever expedient. The equations (\ref {eq:canpi}) above, {\em together}
with the constraint equations (\ref {eq:genCgA}) are now supposed to be solved for the quantities
$\dot{A}_{k}^{a}$ and $C^{a}$ in terms of the canonical coordinates $A_{a}^{k}$ and momenta 
$\pi_{k}^{a}$, respectively. 

Using Eqns. (\ref {eq:genCgA}) and (\ref {eq:Gauss2}), one finds immediately from Eq. 
(\ref {eq:canpi}) that
\begin{equation}
- \nabla_{k}(A)\nabla^{k}(A) C = \nabla^{k}(A)\pi_{k},
\label{eq:defCbc}
\end{equation}
which, together with appropriate boundary conditions at $\partial V$, defines the quantity $C$ as
an $x$-dependent functional of ${\bf A}$ and ${\bf \pi}$, respectively,
\begin{equation}
C  =  C\left \{ {\bf A}, {\bf \pi} \right \}(x^{0}, {\bf x}).
\label{eq:defC2}
\end{equation}
The boundary conditions alluded to above are very important for the {\em consistency} of the Hamiltonian
formalism, as will be seen presently.

One now straightforwardly expresses the generalised velocity in terms of coordinate- and momentum 
variables,
\begin{equation}
\dot{A}_{k}^{a} = \left (\nabla_{k}(A_{0}\left \{{\bf A} \right \} - C) \right )^{a} - \pi_{k}^{a}.
\label{eq:defAdot}
\end{equation}
The construction of the Hamiltonian $H$ then proceeds in the usual way. The relation defining the
Hamiltonian $H$ is the following,
\begin{equation}
H = \int_{V} d^{3}{\bf x} (\pi_{k}, \dot{A}^{k}) - L_{00},
\label{eq:defH}
\end{equation}
where the quantity $\dot{A}^{k}$ ocurring in the expressions in (\ref {eq:defH}) should be 
given in terms of canonical variables by the expression (\ref {eq:defAdot}). It should be observed,
that it is indeed the Lagrangian $L_{00}$ which enters in the definition of the 
Hamiltonian $H$ above, since at this stage the constraint (\ref {eq:genCgA}) is an identity.

By straightforward calculation one finally obtains the Hamiltonian expressed in terms of canonical 
variables from the definition (\ref {eq:defH}),  
\begin{eqnarray}
\label{eq:finHam}
H & = & -\frac{1}{2} \int_{V}d^{3}{\bf x} \left (\pi_{k}, \pi^{k} \right ) + \frac{1}{4} \int_{V} d^{3} {\bf x} \left ( G_{kl}(A), G^{kl}(A) \right )  \\
& & + \int_{V}d^{3}{\bf x} \left (\pi_{k}, \nabla^{k}(A)A^{0}\left \{{\bf A} \right \} \right ) + \frac{1}{2} \int_{V}d^{3}{\bf x} \left (\nabla_{k}(A)C, \nabla^{k}(A)C \right ), \nonumber
\end{eqnarray}
where the quantity $C$ is the appropriate  solution to the system of linear elliptic partial 
differential equations (\ref {eq:defCbc}), as discussed previously. 

By straightforward functional differentiation of the Hamiltonian H in (\ref {eq:finHam}) one finds,
\begin{equation}
\dot{A}_{k}^{a}(x^{0}, {\bf x}) \equiv \frac{\delta H}{\delta \pi_{a}^{k}(x^{0}, {\bf x})} =   \left (\nabla_{k}(A)(A_{0}\left \{{\bf A}\right \} - C)\right )^{a} - \pi_{k}^{a},
\label{eq:dHdpi}
\end{equation}
which agrees precisely with the expression (\ref {eq:defAdot}) as it should. 

It is much more intricate to obtain the second half of the Hamiltonian  equations, which give the 
time evolution of the canonical momentum variables $\pi^{k}_{a}$. It turns out, that the quantity 
$C$ occurring in the expression (\ref {eq:finHam}), which is a solution to the system of 
linear elliptic partial differential equations (\ref {eq:defCbc}), must satisfy a boundary condition 
of the {\em Neumann type} at the boundary $\partial V$ of the domain $V$ in order that the 
functional derivative of the Hamiltonian with respect to the canonical coordinate variables 
$A_{k}^{a}$ be defined. The boundary condition in question is the following,
\begin{equation}
\partial_{n}C|_{{\bf x} \in \partial V}\;= \partial_{n}A_{0}\left \{{\bf A}\right \}|_{{\bf x} \in \partial V},
\label{eq:bcC}
\end{equation}
where $\partial_{n}$ stands for the {\em normal derivative} at the surface $\partial V$, i.e. the
derivative in the direction of the outer normal of $\partial V$. It should be observed, that the 
quantity $A_{0}$ in (\ref {eq:bcC}) is itself a solution to a system of partial differential
equations, namely Eq. (\ref {eq:Gauss2}), for which {\em Dirichlet type} boundary conditions have
to be used, in order to ensure the consistency of the {\em Lagrangian formalism}, as discussed
previously. The role of the boundary conditions in the present context, both Dirichlet- and 
Neumann-, will be considerd in more detail elsewhere \cite{Chris4}.

Under the condition (\ref {eq:bcC}) above, one finds after a lengthy but fairly straightforward 
calculation, that
\begin{eqnarray}
\label{eq:dHdA}
\dot{\pi}_{a}^{k}(x^{0}, {\bf x}) \equiv - \frac{\delta H}{\delta A_{k}^{a}(x^{0}, {\bf x})}& = & -ig[A_{0}\left \{{\bf A}  \right \} - C, \nabla^{k}(A)C + \pi^{k}]_{a}  \\
& & + \left (\nabla_{\ell}G^{k\ell}(A) \right )_{a} - ig[C, \nabla^{k}(A)A_{0}\left \{{\bf A}  \right \}]_{a} \nonumber
\end{eqnarray}

Needless to say, the pair of Hamiltonian equations of motion, Eqns. (\ref {eq:dHdpi}) and
(\ref {eq:dHdA}) are in perfect agreement with the Lagrangian equations of motion obtained from the
variation principle,
\begin{equation}
\delta \int dx^{0} L' = 0,
\label{eq:lastvar}
\end{equation}
where the Lagrangian $L'$ is given in Eq. (\ref {eq:Lconstr}). 

We have thus finally arrived at a canonical Hamiltonian formalism for the Yang-Mills system,
in the case when Gauss' law is supposed to be identically satisfied. As was seen above, this
required, among other things, that a special gauge condition, called the generalised Coulomb
gauge condition \cite{ChrgenCg}, was in force.

The interest in the canonical structure of Yang-Mills theory comes from the interest in quantising 
this theory, either by canonical operator methods or by functional integral methods. I hope to 
return to such questions in the near future.

Before concluding this subsection, it is appropriate to notice, that the canonical Hamiltonian 
equations  (\ref {eq:dHdpi}) and (\ref {eq:dHdA}) admit a {\em first integral}, i.e.
\begin{equation}
\partial_{0}(\nabla_{k}(A) \pi^{k}) = 0.
\label{eq:como}
\end{equation}
The quantity $\nabla_{k}(A)\pi^{k}$ is thus a constant of motion. This is related to the fact, 
that the description of the Yang-Mills system in terms of the canonical variables $A_{a}^{k}$
and $\pi_{k}^{a}$ introduced here, is not a description in terms of a minimal number of dynamical 
degrees of freedom. Also this  circumstance will be discussed in more detail elesewhere 
\cite{Chris4}.

\subsection{Comparison with the Weyl gauge formalism}

For completeness I present below the Hamiltonian formulation of Yang-Mills theory in the 
so-called Weyl gauge \cite{Weyl} $A_{0} = 0$. This has been discussed with admirable clarity
in a paper by Jackiw \cite{Jackiw}, to which I refer for details. This paper also discusses the 
Schr\"{o}dinger quantisation of the Weyl gauge version of Yang-Mills theory. 

The Yang-Mills Lagrangian in the case $A_{0} = 0$, which will be called $L_{W}$ here, is the 
following,
\begin{equation}
L_{W} = - \frac{1}{2} \int d^{3}{\bf x} (\dot{A}_{k}, \dot{A}^{k}) - \frac{1}{4} \int d^{3}{\bf x}(G_{k\ell}, G^{k\ell})
\label{eq:LWeyl}
\end{equation}
The Lagrangian (\ref {eq:LWeyl}) describes a theory which is {\em not} the same as Yang-Mills theory,
since Gauss' law is absent from  the field equations following from the action principle with the
expression  (\ref {eq:LWeyl}) as Lagrangian. Gauss' law takes the following form in the present 
case,
\begin{equation}
\nabla_{k}(A) \dot{A}^{k} = 0
\label{eq:GWeyl}
\end{equation}
The traditional way to analyse Yang-Mills theory in the Weyl gauge, is to proceed from the Lagrangian
(\ref {eq:LWeyl}), disregarding Gauss' law to begin with.

Using the variables $A_{k}^{a}$ and $\dot{A}_{k}^{a}$ as generalised coordinates and velocities, 
respectively, it is perfectly simple to derive a canonical Hamiltonian formalism for the system
defined by the Lagrangian (\ref {eq:LWeyl}). The corresponding Hamiltonian $H_{W}$ is,
\begin{equation}
H_{W} =  - \frac{1}{2} \int d^{3}{\bf x} (\pi_{k}, \pi^{k}) + \frac{1}{4} \int d^{3}{\bf x}(G_{k\ell}, G^{k\ell})
\label{eq:HamWeyl}    
\end{equation}
i.e. a simple sum of a kinetic term, depending on $\pi$ only, and a potential, or interaction term, 
depending on $A$ only.

The Hamiltonian $H_{W}$ in Eq. (\ref {eq:HamWeyl}) should now be compared with the Hamiltonian 
(\ref {eq:finHam}) derived in the previous subsection. Disregarding for the time being the 
slightly different meaning of the symbols in the two Hamiltonians in question, one observes that 
the Hamiltonian (\ref {eq:finHam}) contains interaction terms depending on the quantities $A_{0}$ 
and $C$ respectively, which are completely absent from the Weyl gauge Hamiltonian (\ref {eq:HamWeyl}).

A transition to quantum versions of the Hamiltonians above, by means of the fixed time 
Schr\"{o}dinger quantisation rule,
\begin{equation}
\pi_{k}^{a} \rightarrow \hat{\pi}_{k}^{a} \equiv -i\frac{\delta}{\delta A_{a}^{k}}
\label{eq:QuSchr}
\end{equation}
facilitates a comparison of the two Hamiltonians. 

Interpreting the Hamiltonian (\ref {eq:finHam}) as a quantum operator $\hat{H}$ by means of the 
substitution (\ref {eq:QuSchr}) one is then invited to consider the following eigenvalue equation, 
in self-explanatory notation,
\begin{equation}
\hat{H}\Psi(A) = E\Psi(A)     
\label{eq:eigH}
\end{equation}
which in principle should describe a glueball spectrum. There are pitfalls associated 
with the quantisation procedure briefly described above, which have to be understood and avoided 
before considering conclusions based on equations such as (\ref {eq:eigH}), as being ironclad.

In the Weyl gauge case one would have, similarily,
\begin{equation}
\hat{H}_{W}\Psi_{W}(A) = E\Psi_{W}(A)     
\label{eq:eigHW}
\end{equation}

At this stage one simply adds the hitherto omitted Gauss' law (\ref {eq:GWeyl}) as a condition 
on the states $\Psi_{W}(A)$,
\begin{equation}
\left ( \frac{\partial}{\partial x^{k}}\frac{\delta}{\delta A_{k}^{a}(x^{0}, {\bf x})} -gf_{ab}^{~~c}A_{k}^{b}(x^{0}, {\bf x})\frac{\delta}{\delta A_{k}^{c}(x^{0}, {\bf x})} \right )\Psi_{W}(A) = 0
\label{eq:GaWstates}
\end{equation}

It is now appropriate to recall the first integral (\ref {eq:como}) admitted by the 
(semiclassical) system described by the Hamiltonian $H$ in (\ref {eq:finHam}). In the 
Schr\"{o}dinger quantisation of that system, the relation (\ref {eq:como}) gives rise to a 
condition on the states $\Psi(A)$  of the same type as (\ref {eq:GaWstates}), although not 
necessarily identical to that relation. Even if one might reproduce exactly a condition
of the form (\ref {eq:GaWstates}) in the Schr\"{o}dinger quantised theory defined by the quantum
Hamiltonian $\hat{H}$, by an appropriate choice of a constant of integration, one must still 
reckon with the fact that the Hamiltonian $\hat{H}$ definitely differs from the Hamiltonian 
$\hat{H}_{W}$ by the interaction terms involving the quantities $A_{0}$ and $C$ in the 
(semiclassical) expression (\ref {eq:finHam}). These interaction terms, appeared as a result of 
the demand that Gauss' law should be satisfied identically. The variables $A_{0}$ and $C$ in 
these intreraction terms had to satisfy certain very precise boundary conditions
in order to ensure the consistency of the (semiclassical) Lagrangian and Hamiltonian formulation, as
explained previously. Hence these interaction terms cannot be declared to be uneccesary or 
unmiportant. So, one is forced to conclude that the Weyl gauge quantum canonical formalism 
in Yang-Mills theory differs in substance from the corresponding formalism developed in this paper.

\section{Summary and Conclusions}   

In this paper I have considered the problem of defining canonical coordinates and momenta in pure 
Yang-Mills theory, under the condition that Gauss' law is identically satisifed. Satisfying Gauss'
law in this context means solving a system of  elliptic linear partial differential equations for 
the matrix valued gauge potential component $A_{0}$, with a fixed value of the time coordinate. For
a discussion of the canonical structure of Yang-Mills theory, when Gauss' law is identically 
satisifed, one does not need the explicit form of the solutions to the differential equations in
question, but merely certain properties of the solutions, which are embodied in the equations 
themselves. Existence of solutions is not proved here, but assumed. The related question of 
uniqueness of solutions, which depends on boundary conditions, is only briefly touched upon. 
The requisite boundary conditions for the system of equations are not postulated {\em a priori}, but 
arise as consistency conditions related to the equations of motion. The canonical structure of the 
theory requires the choice of a special gauge, called the generalised Coulomb gauge in this paper. 
This gauge, the existence of which is equivalent to the existence of appropriate solutions to
a system of partial differential equations of the same type as encountered in connection with Gauss'
law, does in principle not suffer from Gribov ambiguitites. 

The canonical variables, coordinates and momenta, as well as the Hamiltonian, are explicitly 
constructed. This construction involves again a system of partial differential equations of the 
same type as encountered previously, but now with different boundary conditions, which are dictated 
by necessary consistency conditions for the existence of canonical Hamiltonian equations of motion.

Schr\"{o}dinger quantisation of the theory is briefly discussed, and compared with the corresponding
quantum theory in the Weyl gauge, $A_{0} = 0$. It is concluded that the Schr\"{o}dinger quantised 
Hamiltonian formulation of Yang-Mills theory developed in this paper, differs in substance from the 
corresponding Weyl gauge formulation.

\vspace*{1cm} 
\noindent 
{\bf Acknowledgements}

I am indebted to my colleague Claus Montonen for useful comments on the manuscript of this paper.

\end{document}